\documentclass[
 reprint,
 amsmath,amssymb,
 aps,
]{revtex4-2}

\usepackage{graphicx}
\usepackage{hyperref}
\usepackage[linesnumbered, ruled, vlined, titlenotnumbered, noend]{algorithm2e}
\usepackage{amsmath, amsfonts, amssymb, amsthm, color}
\usepackage{tikz-cd}
\usepackage[capitalise]{cleveref}
\crefname{algocf}{alg.}{algs.}
\Crefname{algocf}{Algorithm}{Algorithms}

\newcommand{\bfA}{{\mathbf{A}}}
\newcommand{\bfH}{{\mathbf{H}}}
\newcommand{\bfp}{{\mathbf{p}}}
\newcommand{\bfe}{{\mathbf{e}}}
\newcommand{\bfs}{{\mathbf{s}}}
\newcommand{\F}{{\mathbb{F}}}

\newcommand{\wt}{\text{wt}}

\newcommand{\sign}{\text{sign}}

\SetKwInOut{KwParams}{Parameters}
\SetKwProg{Func}{Function}{}

\begin{document}

\title{Beam search decoder for quantum LDPC codes}
\author{Min Ye, Dave Wecker, Nicolas Delfosse}
\affiliation{
    IonQ Inc.
}

\date{\today}

\begin{abstract}
We propose a decoder for quantum low density parity check (LDPC) codes based on a beam search heuristic guided by belief propagation (BP).
Our beam search decoder applies to all quantum LDPC codes and achieves different speed-accuracy tradeoffs by tuning its parameters such as the beam width.
We perform numerical simulations under circuit level noise for the $[[144, 12, 12]]$ bivariate bicycle (BB) code at noise rate $p=10^{-3}$ to estimate the logical error rate and the 99.9 percentile runtime and we compare with the BP-OSD decoder which has been the default quantum LDPC decoder for the past six years.
A variant of our beam search decoder with a beam width of 64 achieves a $17\times$ reduction in logical error rate.
With a beam width of 8, we reach the same logical error rate as BP-OSD with a $26.2\times$ reduction in the 99.9 percentile runtime.
We identify the beam search decoder with beam width of 32 as a promising candidate for trapped ion architectures because it achieves a $5.6\times$ reduction in logical error rate with a 99.9 percentile runtime per syndrome extraction round below 1ms at $p=5 \times10^{-4}$.
Remarkably, this is achieved in software on a single core, without any parallelization or specialized hardware (FPGA, ASIC), suggesting one might only need three 32-core CPUs to decode a trapped ion quantum computer with 1000 logical qubits.
\end{abstract}

\maketitle

\section{Introduction}

Classical low-density parity-check (LDPC) codes~\cite{gallager1962low, mackay2003information, richardson2008modern} are widely adopted in classical information processing (WiFi, 5G mobile, and flash memory).
One of the main reasons for the success of LDPC codes is that they come with a fast and accurate decoder, the so-called belief propagation (BP) decoder~\cite{pearl1982reverend}.

The generalization of LDPC codes to the quantum setting was originally proposed by Mackay, Mitchison and McFadden~\cite{mackay2004sparse}. 
Later several more efficient constructions of quantum LDPC codes were introduced, such as hypergraph product (HGP) codes~\cite{tillich2013quantum}, two-block codes~\cite{kovalev2013quantum}, balanced product codes~\cite{breuckmann2021balanced},
and the recently discovered asymptotically good quantum LDPC codes~\cite{panteleev2022asymptotically, leverrier2022quantum}.
Moreover, circuit level simulations proved that several instances of quantum LDPC codes outperform surface codes such as hyperbolic codes~\cite{higgott2024constructions}, bivariate bicycle (BB) codes~\cite{bravyi2024high} or their BB5 variant~\cite{ye2025quantum}, radial code~\cite{scruby2024high}, or HGP codes~\cite{tremblay2022constant, aydin2025cyclic}.
Precisely, they achieve the same logical error rate as surface codes with a substantially smaller qubit overhead.

However, to make quantum LDPC codes practical, a fast and accurate decoder is needed.
BP is fast but it does not perform well in general when applied to quantum LDPC codes.
The first issue comes from the definition of the decoding problem.
In the classical case, BP is designed to estimate the marginal error probability of each bit.
However, the concept of marginal error probability for a single qubit in a stabilizer code is not well-defined. Since any error is equivalent to its product with a stabilizer, an error acting non-trivially on a specific qubit is equivalent to another error that acts trivially on that same qubit~\cite{gottesman1997stabilizer}.
The second issue is that BP often fails to converge to a valid correction.
To explain this problem, recall that BP works as a message-passing algorithm, sending data through the edges of the graph representing the code, which we call the Tanner graph~\cite{tanner1981recursive}.
For BP to be accurate, the code must be designed in such a way that its Tanner graph is cycle-free, so that a message sent from a node cannot loop back to its sender, avoiding risks of inconsistencies.
For a non-trivial code, it is impossible to remove all cycles but removing short cycles is enough to ensure a high accuracy for BP~\cite{hu2001progressive}.
Unfortunately, the structure of quantum LDPC codes makes short cycles unavoidable~\cite{mackay2004sparse}.
As a result, some marginal probabilities oscillate and BP either fails to converge, reducing its accuracy, or BP converges slowly, reducing its speed.

The BP-OSD decoder, proposed in 2019 by Panteleev and Kalachev~\cite{panteleev2021degenerate} and improved in~\cite{roffe_decoding_2020, Roffe_LDPC_Python_tools_2022}, has been the default decoder for quantum LDPC code simulations over the past six years.
It is far more accurate than BP, achieving logical error rates orders of magnitude better.
However, it is too slow for practical applications because it relies on a matrix inversion implemented in cubic complexity by Gaussian elimination.
This matrix inversion is also the bottleneck when adapting the union find decoder to quantum LDPC codes~\cite{delfosse2022toward}.

The ambiguity clustering (AC) decoder~\cite{wolanski2024introducing}
and the BP-LSD decoder~\cite{hillmann2024localized} eliminate the BP-OSD bottleneck by partitioning the matrix inversion into sub-problems that can be resolved independently, effectively reducing the average runtime, but these approaches do not improve the accuracy of BP-OSD. 
Moreover, the worst-case runtime of these decoders remains superlinear, which impacts the tail of the runtime distribution. For example, in Fig.~9 of \cite{hillmann2024localized}, we see that the tail of the runtime distribution of BP-LSD reaches runtimes that are two orders of magnitude larger than the average.
To avoid this issue, which could lead to a large qubit and time overhead for fault-tolerant quantum computation~\cite{terhal2015quantum, khalid2025impacts}, we track both the average runtime and the 99.9 percentile runtime of our decoder.

\begin{figure*}
    \centering
    \includegraphics[width=.95\linewidth]{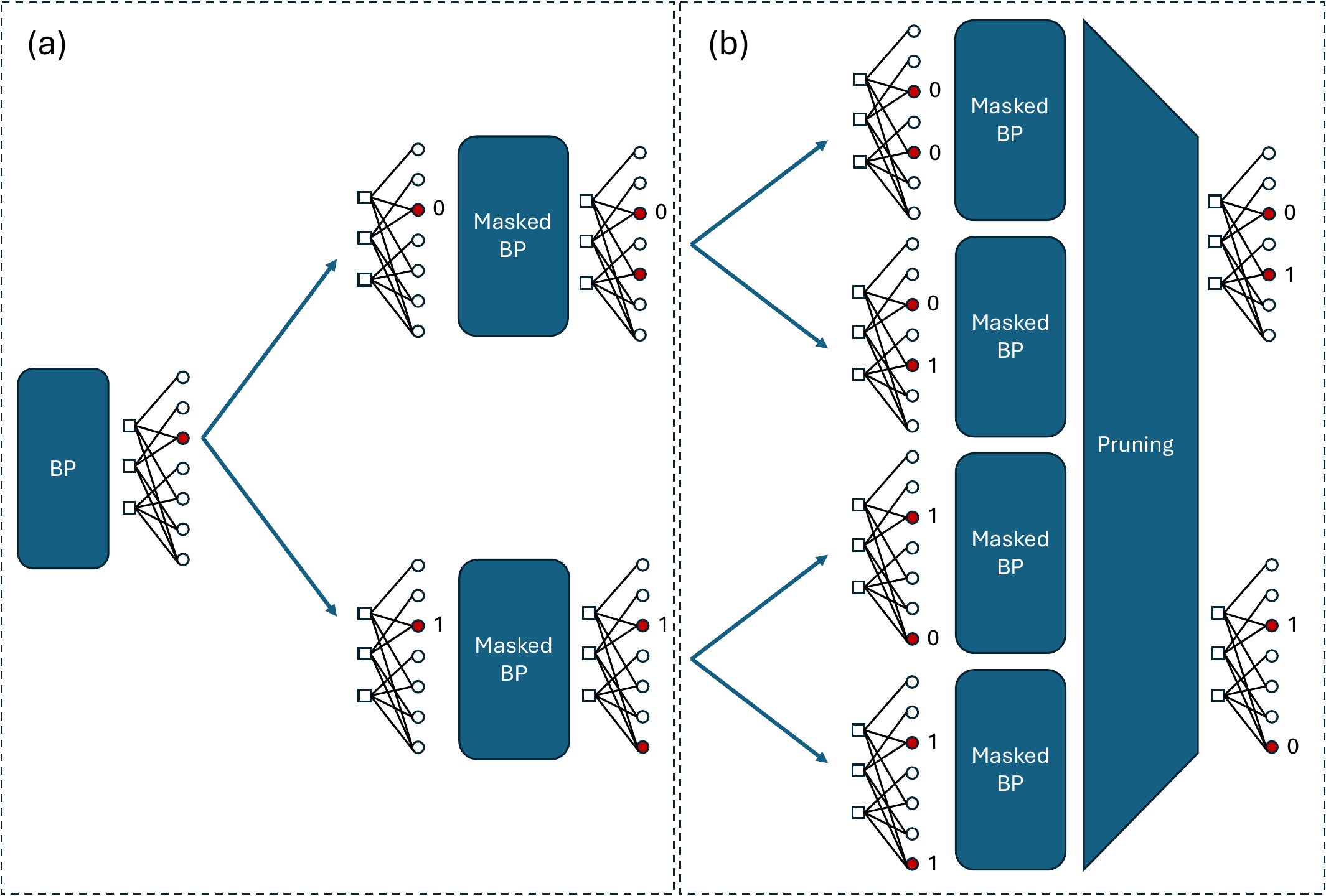}
    \caption{Overview of the beam search decoder. 
    (a) The decoder is initialized with a small number of BP iterations.
    Then, the least reliable error node (red) is selected and two branches are created corresponding to the two possible values of this node.
    After the first step, BP is replaced by a masked BP ignoring the previously fixed error nodes.
    (b)
    The beam search decoder repeats the following three steps: (i) branching over the least reliable error node (ii) running a masked BP and (iii) pruning to reduce the number of paths to the beam width (which is 2 in this figure).
    The decoder is terminated once a sufficient number of solutions is found or if a maximum number of repetitions is reached.
    }
    \label{fig:beam_search_steps}
\end{figure*}

Another strategy to improve BP for quantum LDPC codes is to modify the message-passing procedure by adding or removing constraints in order to suppress the impact of short cycles and the marginal oscillations.
Recall that BP sends messages back and forth between error nodes, representing potential error sources, and detector nodes associated with syndrome measurement results.
Poulin and Chung proposed to freeze the value of an error node neighboring an unsatisfied detector node in the Tanner graph~\cite{poulin2008iterative}.
Stabilizer inactivation (SI) removes the value of the least reliable stabilizer node~\cite{du2022stabilizer}.
Guided decimation (BP-GD) fixes the value of the most reliable qubit node~\cite{yao2024belief, gong2024toward}.
Guided decimation is combined with backtracking in the decision-tree decoder of~\cite{ott2025decision}.
The ordered Tanner forest post-processing (BP-OTF) is executed in a cycle-free subgraph of the Tanner graph~\cite{iolius2024almost}.
BP-Relay introduces a memory in BP~\cite{muller2025improved}.
Remarkably, SI, BP-GD, the decision-tree decoder and BP-Relay achieve a better accuracy than BP-OSD for the $[[144, 12, 12]]$ BB code with circuit level noise.
The runtime distribution tail was not analyzed in these papers.
When implemented on FPGA, the runtime of BP Relay seems promising for applications to superconducting qubits but the introduction of a long-term memory makes it inherently sequential which might lead to a long runtime tail at the software level~\cite{maurer2025real}.

Different parallel variants of BP were proposed.
The automorphism-based decoder of~\cite{koutsioumpas2025automorphism} runs BP in parallel over many inputs obtained by applying an automorphism of the code.
The parallel BP decoder of~\cite{wang2025fully} generates several initial configurations by flipping some of the most oscillating qubits and runs BP in parallel for all these initial configurations.
These decoders are embarrassingly parallel but they do not significantly improve the accuracy of BP-OSD at the circuit level.

Optimization algorithms have also been used to design decoders for quantum LDPC codes.
The decision tree decoder of \cite{ott2025decision} explores the tree of all fault-configurations starting from the empty set.
Wu {\em et al.} treated the decoding problem as a linear program and solved it together with its dual to obtain performance guarantees~\cite{wu2025minimum}.
An integer programming decoder and a decoder based on A* search were considered in~\cite{beni2025tesseract}.
These four decoders can be useful to probe the optimal performance of small codes but they are too slow for real-time decoding.
A recent branch and bound decoder might be faster but it was not analyzed under circuit level noise~\cite{valentini2025restart}.

In this work, we propose a beam search decoder which is simultaneously fast, accurate, easy to parallelize, and flexible.
The decoder is parametrized by the beam width and other parameters.
We estimate the performance of our beam search decoder with numerical simulations for the $[[144,12,12]]$ bivariate bicycle (BB) code under circuit level noise with noise rate $p=10^{-3}$.
The simulations are performed on a 2023 M3 processor on a single core without any parallelization.
Our beam search decoder achieves a logical error rate that is up to $17\times$ better than BP-OSD.
For a beam width of 8, we achieve the same logical error rate, a $4.6 \times$ reduction of the average runtime and a $26.2\times$ reduction of the 99.9 percentile runtime compared with BP-OSD.
For a beam width of 32, we achieve a $5.6 \times$ reduction of the logical error rate, a $2.8 \times$ reduction of the average runtime and a $20.4\times$ reduction of the 99.9 percentile runtime compared with BP-OSD.
This proves that the beam search decoder can be simultaneously more accurate and faster than BP-OSD.

To examine the practical application of our beam search decoder in a large-scale fault-tolerant trapped ion quantum computer, we consider its performance at lower noise rate. 
We pick $p=5 \times 10^{-4}$, which is above the noise rate achievable in today's devices~\cite{hughes2025trapped}, and our goal is to design a high-accuracy decoder whose 99.9 percentile runtime per syndrome extraction round is under 1ms, which is the expected syndrome extraction time on trapped ion machines.
At this noise rate and with the $[[144, 12, 12]]$ BB code, the beam search decoder with beam width of 32 has an average runtime per syndrome extraction round of $270 \mu s$ and a 99.9 percentile runtime of $940 \mu s$, which is 24 times better than BP-OSD.

Our work proves that a software-level decoder on a single core, without any parallelization or specialized hardware (FPGA, ASIC) can simultaneously achieve a significantly better logical error rate than BP-OSD and a 99.9 percentile runtime below 1ms.
This suggests that one might only need three 32-core CPUs to decode a trapped ion quantum computer with 1000 logical qubits. 
For comparison, a fault-tolerant quantum computing architecture based on superconducting qubits and surface codes might need up to 1000 ASICs or FPGA decoders to correct 1000 surface code patches simultaneously.

We further demonstrate the flexibility of our decoder by performing numerical simulations showing that the beam search decoder also outperforms BP-OSD for other BB codes and for HGP codes~\cite{aydin2025cyclic}.
Moreover, it can benefit from the XYZ-decoding, which utilizes both X and Z syndrome outcomes simultaneously for decoding~\cite{maurer2025real}.

The rest of this paper is organized as follows. \cref{sec:overview} introduces the main ideas of the beam search decoder. In \cref{sec:simulations}, we present our numerical results, followed by concluding remarks in \cref{sec:conclusion}. Additionally, Appendix~\ref{sect:background} reviews the quantum decoding problem and the BP algorithm, while Appendix~\ref{sec:beam_search_decoder} provides a detailed algorithmic description of the beam search decoder.

\begin{algorithm}
    \DontPrintSemicolon
    \SetAlgoLined
    Run standard BP and return if it converges.\;
    $\texttt{path.next\_pos} \gets$ the least reliable error node in BP.\;
    $\texttt{path.pos\_val\_pairs} \gets$ an empty vector.\;
    Initialize \texttt{set} $\gets\{$\texttt{path}$\}$ with a single element \texttt{path}\;
    \For{$r=1,2,\dots,$\emph{\texttt{max\_rounds}}}{
    Initialize \texttt{next\_set} as an empty set.\;
    \For{\emph{each \texttt{path}$\in$\texttt{set} and \texttt{val}}$\in \{0,1\}$}{
    $\texttt{nextp.pos\_val\_pairs} \gets \texttt{path.pos\_val\_pairs}\cup (\texttt{path.next\_pos},\texttt{val})$\;
    Run masked BP masking the nodes in \texttt{nextp.pos\_val\_pairs}.\; 
    If masked BP converges, return the result.\;
    $\texttt{nextp.next\_pos} \gets$ the least reliable error node in masked BP.\;
    $\texttt{nextp.score} \gets$ reliability score of \texttt{nextp}.\;
    Add \texttt{nextp} into \texttt{next\_set}.\; 
    Remove the element with the smallest \texttt{score} if the size of \texttt{next\_set} exceeds the parameter \texttt{beam\_width}.\;
    }
    \texttt{set} $\gets$ \texttt{next\_set} \;
    }
    \caption{beam search decoder (high-level sketch)}
    \label{algorithm:sketch_beam_search}
\end{algorithm}

\section{Overview of the beam search decoder} \label{sec:overview}

This section gives an overview of the beam search decoder. The algorithm is initialized by a standard BP run to generate a single ``seed" path. This path is used to form the initial set, which contains only one entry at this stage. This initial run also identifies the least reliable error node, which will be used for the first branching step.

The decoder then executes multiple rounds of ``masked BP". In each round, every path in the set is expanded into two new branches by constraining the least reliable error node (identified in the previous round) to 0 and 1, respectively. This newly constrained node is now considered ``masked", meaning that it is effectively removed from the BP calculations. This constraint is added to the path's history of masked error nodes. To maintain tractable complexity, this branching step is followed by a pruning step: the set is sorted by a reliability score, and only a fixed number of the most reliable paths are kept. This process repeats until one of two termination conditions is met: either a pre-defined number of valid solutions is collected, or the maximum number of masked BP rounds is reached. The decoder then returns the minimum-weight solution it finds. \cref{fig:beam_search_steps} illustrates the workflow of the decoding algorithm, while \cref{algorithm:sketch_beam_search} presents a high-level sketch for the special case where the decoder terminates upon finding the first valid solution.

In our decoder, the reliability metric for each error node is the absolute value of the sum of its posterior LLRs over all BP iterations in the current round. We use this sum-based metric rather than the LLR from a single BP iteration to ensure that oscillating error nodes are correctly identified as unreliable. The reliability score of a path, which is used to prune the set, is then calculated by summing these individual node metrics over all of its associated unmasked error nodes.

Pseudocode and a detailed description of the beam search decoder are provided in Appendix~\ref{sec:beam_search_decoder}, and the implementation source code is available on GitHub~\cite{beam_search_github}.

Our beam search decoder shares several similarities with the BP-GDG decoder proposed in \cite{gong2024toward}. Both algorithms, for example, employ multiple rounds of masked BP and select one error node to fix (or constrain) in each round. The BP-GDG decoder, in its initial 4 rounds, also branches by exploring both values (0 and 1) for the selected error node. However, there are three important differences between our decoder and BP-GDG.
First, our decoder systematically branches in every round, whereas BP-GDG only branches for the first 4 rounds. After that, BP-GDG only performs a quick exploration of immediate side branches of the most likely decoding path. 
Second, and most importantly, we introduce a reliability score to predict a path's likelihood of success before its completion. To our knowledge, this is a new approach. In contrast, standard decoders (including BP-GDG) can only evaluate a path's quality after it terminates—either by finding a valid solution and checking its weight, or by failing to converge at the maximum iteration limit. This conventional method must run all paths to completion, even those that are unlikely to succeed. Our reliability score enables the decoder to proactively prune the set, eliminating unpromising paths early and focusing computational effort on more viable candidates, as demonstrated by our simulation results.
Third, the node selection strategies differ. Our decoder branches on the least reliable error node (the one with the minimum absolute summed LLR). In contrast, BP-GDG branches on the error node most likely to be 1 (the one with the most negative LLR history).

\begin{table*}
    \centering
    \begin{tabular}{|c|c|c|c|c|c|}
    \hline
       name  &  \texttt{max\_rounds} & \texttt{beam\_width} & \texttt{initial\_iters} & \texttt{iters\_per\_round} & \texttt{num\_results}  \\ 
       \hline
       \texttt{beam8\_230iters}  & 10 & 8 & 30 & 20 & 1 \\
       \hline
       \texttt{beam32\_340iters} & 10 & 32 & 40 & 30 & 1 \\
       \hline
       \texttt{beam64\_640iters} & 20 & 64 & 40 & 30 & 1 \\
       \hline
       \texttt{beam64\_32res\_640iters} & 20 & 64 & 40 & 30 & 32 \\
       \hline
    \end{tabular}
    \caption{Name and parameters for different configurations of beam search decoder. The number of iterations in name is calculated as $\texttt{initial\_iters}+\texttt{max\_rounds}*\texttt{iters\_per\_round}$. The parameter \texttt{num\_results} is omitted from the name if it is equal to 1.}
    \label{tab:beam_search_decoder_parameters}
\end{table*}

\begin{table*}
    \centering
    \begin{tabular}{|c|c|c|c|c|}
    \hline
      physical error rate & \texttt{bp30+osd}   & \texttt{beam8\_230iters} & \texttt{beam32\_340iters} & \texttt{beam64\_640iters} \\
       \hline
       $5\times 10^{-4}$ & 3.55ms & 1.627ms & 3.202ms & 5.881ms \\
       \hline
       $10^{-3}$ & 10.59ms & 2.318ms & 3.837ms & 6.698ms  \\
       \hline
    \end{tabular}
    \caption{Average decoding time in circuit-level noise simulations of the $[[144,12,12]]$ BB code with 12 syndrome extraction rounds.}
    \label{tab:average_decoding_time}
\end{table*}

\begin{table*}
    \centering
    \begin{tabular}{|c|c|c|c|c|}
    \hline
      physical error & \texttt{bp30+osd}   & \texttt{beam8\_230iters} & \texttt{beam32\_340iters} & \texttt{beam64\_640iters} \\
       \hline
       $5\times 10^{-4}$ & 272.5ms & 8.704ms & 11.26ms & 14.53ms\\
       \hline
       $10^{-3}$ & 289.0ms & 11.01ms & 14.18ms & 17.70ms \\
       \hline
    \end{tabular}
    \caption{The 99.9 percentile decoding time in circuit-level noise simulations of the $[[144,12,12]]$ BB code with 12 syndrome extraction rounds.}
    \label{tab:999percentile_decoding_time}
\end{table*}

\begin{figure}
    \centering
    \includegraphics[width=\linewidth]{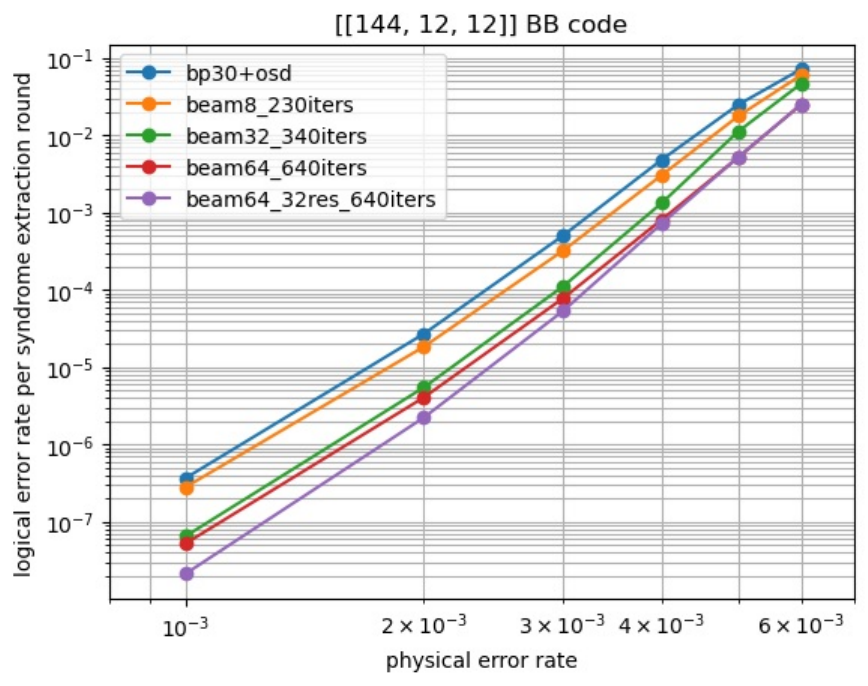}
    \caption{Simulation results for the $[[144,12,12]]$ BB code under circuit-level noise. BP-OSD decoder is configured with 30 min-sum BP iterations followed by order-10 combination-sweep OSD.}
    \label{fig:logical_error_rate_144_12}
\end{figure}

\begin{figure}
    \centering
    \includegraphics[width=\linewidth]{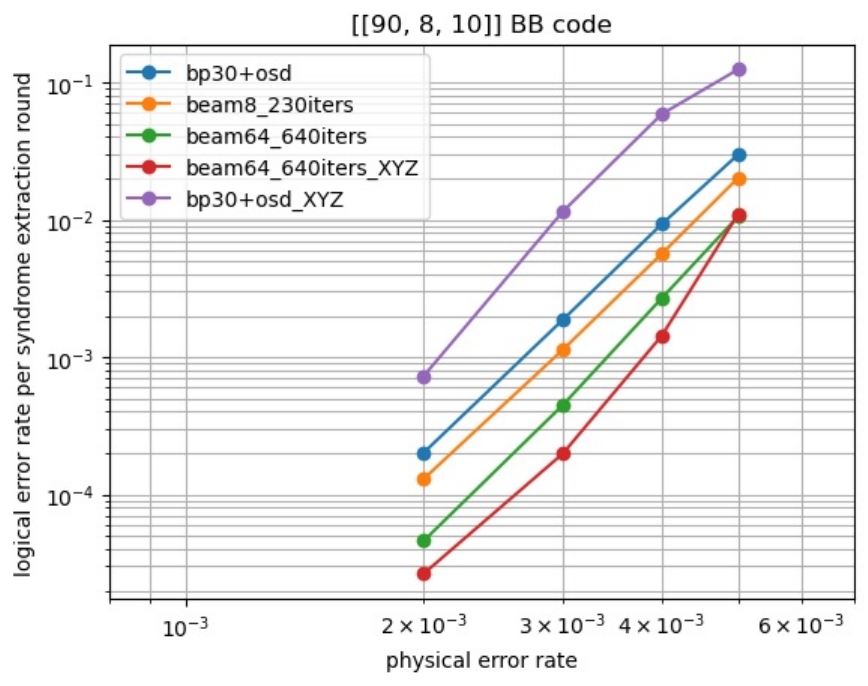}
    \caption{Simulation results for the $[[90,8,10]]$ BB code under circuit-level noise. The suffix \texttt{\_XYZ} in the legend denotes XYZ-decoding, which utilizes both X and Z syndrome outcomes simultaneously for decoding.}
    \label{fig:logical_error_rate_90_8}
\end{figure}

\begin{figure}
    \centering
    \includegraphics[width=\linewidth]{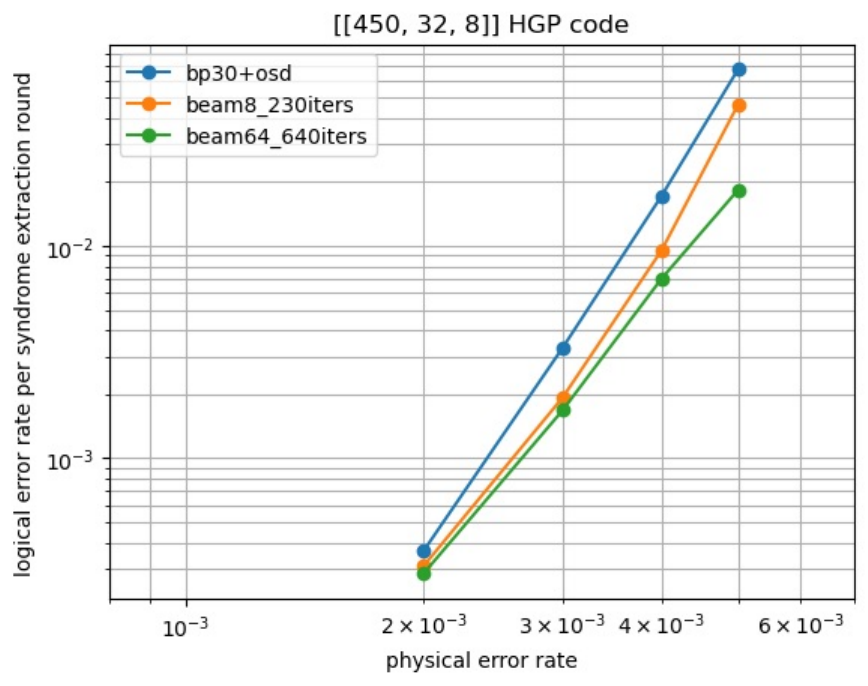}
    \caption{Simulation results for the $[[450,32,8]]$ HGP code \cite{aydin2025cyclic}.}
    \label{fig:logical_error_rate_450_32}
\end{figure}

\section{Simulation results} \label{sec:simulations}

We provide simulation results for several configurations of our beam search decoder across three quantum LDPC codes. We compare the performance of these configurations against a baseline BP-OSD decoder~\cite{panteleev2021degenerate, roffe_decoding_2020, Roffe_LDPC_Python_tools_2022}. Following~\cite{hillmann2024localized}, which compares BP-OSD and BP-LSD, we configure the baseline BP-OSD decoder with 30 min-sum BP iterations and order-10 combination-sweep OSD post-processing.
\cref{tab:beam_search_decoder_parameters} lists the names and corresponding parameters for each beam search decoder configuration used in our simulations.

\cref{fig:logical_error_rate_144_12} plots the logical error rate of the $[[144,12,12]]$ BB code under circuit-level noise. At physical error rate $p=10^{-3}$, even the most computationally efficient configuration, \texttt{beam8\_230iters}, yields a $1.3\times$ improvement over \texttt{bp30+osd}. Increasing the beam width further enhances performance: \texttt{beam32\_340iters} and \texttt{beam64\_640iters} reduce the logical error rate by factors of $5.6\times$ and $7.0\times$, respectively. Notably, the most advanced configuration, \texttt{beam64\_32res\_640iters}, achieves a $17\times$ improvement over BP-OSD.

\cref{tab:average_decoding_time} and \cref{tab:999percentile_decoding_time} detail the decoding time statistics for the $[[144,12,12]]$ BB code over 12 syndrome extraction rounds, measured on a single core of a 2023 Apple M3 Pro. \cref{tab:average_decoding_time} reports the average decoding time, while \cref{tab:999percentile_decoding_time} lists the 99.9 percentile. We evaluate two physical error rates: $p=10^{-3}$, relevant for superconducting qubits, and $p=5\times 10^{-4}$, applicable to higher-fidelity hardware such as trapped ion systems. At both error rates, all three beam search configurations improve the 99.9 percentile decoding time by at least $16\times$ compared to the \texttt{bp30+osd} baseline, demonstrating superior capability in mitigating worst-case decoding latency. Specifically, the fastest configuration, \texttt{beam8\_230iters}, achieves reductions of $31\times$ and $26\times$ at $p=5\times 10^{-4}$ and $p=10^{-3}$, respectively. Benefiting from the significantly reduced variance in decoding time, the \texttt{beam32\_340iters} configuration achieves a 99.9 percentile time of less than 1ms per syndrome extraction round at $p=5\times 10^{-4}$. While the reduction in average time is less pronounced, \texttt{beam8\_230iters} still yields a $4.6\times$ speedup at $p=10^{-3}$.

\cref{fig:logical_error_rate_90_8} plots the logical error rate of the $[[90,8,10]]$ BB code using two strategies: standard XZ-decoding and XYZ-decoding, following the terminology in~\cite{muller2025improved}. Since both BB codes and HGP codes simulated in this paper are CSS codes, X and Z errors are decoded separately, and the total logical error rate is approximated as the sum of the X and Z logical error rates. Recall from \cref{sect:background} that the decoding problem is defined by the triple $(\bfH, \bfA, \bfp)$. Both strategies utilize the same logical operator matrix $\bfA$ and error probabilities $\bfp$; they differ only in the construction of the parity-check matrix $\bfH$. When decoding X errors, both strategies employ the same logical operator matrix $\bfA$ consisting of Z logical operators; however, XZ-decoding constructs $\bfH$ using only Z stabilizers, whereas XYZ-decoding includes both X- and Z-type stabilizers. Similarly, when decoding Z errors, XZ-decoding uses only X stabilizers to form $\bfH$, while XYZ-decoding employs the full set of stabilizers.

In theory, utilizing the full set of stabilizers yields more information than restricting the decoder to a single stabilizer type. This advantage arises because the circuit-level noise model assumes that qubits undergo depolarizing noise during unitary gates and idling time. Since depolarizing noise includes a Pauli-Y component, which triggers both X- and Z-type stabilizers, employing both types simultaneously captures strictly more information.

However, a major drawback of utilizing the full set of stabilizers is the induction of length-4 cycles in the corresponding Tanner graph, which is known to degrade the performance of BP and many BP-based decoders. As shown in \cref{fig:logical_error_rate_90_8}, the \texttt{bp30+osd} decoder performs much worse under XYZ-decoding than XZ-decoding, despite the fact that XYZ-decoding theoretically provides more information. In contrast, \texttt{beam64\_640iters} effectively leverages this additional information, reducing the logical error rate by approximately $2\times$ compared to XZ-decoding in the range $0.002\le p\le 0.004$. A likely explanation is that the short cycles in XYZ-decoding hinder BP-based decoders by causing the posterior LLRs of error nodes to oscillate between positive and negative values. However, because the reliability score in our beam search decoder is derived from the absolute sum of posterior LLRs, it can successfully identify these oscillating nodes as unreliable and exclude them from subsequent BP iterations.

Finally, to verify the robustness of our decoder across different quantum LDPC families, we perform simulations of the $[[450,32,8]]$ cyclic HGP code introduced in~\cite{aydin2025cyclic}. The logical error rates for three decoders are presented in \cref{fig:logical_error_rate_450_32}. Our beam search decoders continue to exhibit lower logical error rates than the \texttt{bp30+osd} baseline. In particular, at a relatively high noise rate $p=0.005$, the \texttt{beam8\_230iters} and \texttt{beam64\_640iters} configurations achieve improvements of $1.4\times$ and $3.7\times$, respectively.

\section{Conclusion} \label{sec:conclusion}

In this work, we propose a simple beam search decoder for quantum LDPC codes. We show that it is simultaneously fast and accurate, significantly outperforming the BP-OSD decoder.

We demonstrate that the beam search decoder with beam width of 32 achieves a lower logical error rate than BP-OSD while satisfying the runtime requirements for trapped ion or neutral atom quantum computers, which is expected to be of the order of 1ms. Notably, this configuration meets strict latency constraints, achieving a sub-millisecond runtime not only on average but also at the 99.9 percentile.

It is a folklore that superconducting qubits, which must be decoded within a micro-second window, might require a supercomputer to perform decoding for a large-scale fault-tolerant quantum computer.
This encouraged researchers to optimize decoders at the micro-architecture level~\cite{das2022afs} and to build hardware-level decoders on FPGAs or ASICs~\cite{liyanage2023scalable, liyanage2024fpga, caune2024demonstrating, barber2025real, ziad2024local, wu2025micro, valentino2025quekuf, maurer2025real, maurya2025fpga}.
In contrast, because the decoding time budget of trapped ions and neutral quantum computer is significantly larger, our decoder can be executed on a CPU without any parallelization or specialized hardware implementation.

The BP algorithm appeared independently in different communities. It is used in coding theory, statistical physics, machine learning and optimization and it led to fruitful exchanges of ideas between these communities~\cite{nishimori2001statistical, mezard2009information, mackay2003information, richardson2008modern}.
In the quantum setting, the interface between quantum error correction and statistical physics, is an active research topic~\cite{dennis2002topological, delfosse2010quantum, bombin2012strong, kovalev2013spin, kubica2018three, li2021statistical, vodola2022fundamental, placke2024topological, placke2025expansion, english2025thresholds}.
We anticipate that the beam search decoder will play a role at this interface, similar to the connecting role of BP in the classical case.

Beyond its practical utility, the simplicity of the beam search decoder makes it a promising candidate for theoretical analysis. We hope this feature will facilitate the proof of asymptotic results for quantum LDPC codes, much like the BP decoder proved crucial in the design of capacity-achieving classical LDPC codes~\cite{richardson2008modern}.

\vspace*{0.1in}
\section*{Acknowledgment}

The authors thank Aharon Brodutch, Edwin Tham, Felix Tripier, Joe Latone and John Gamble and the whole IonQ team for insightful discussions.




\bibliography{references}

\appendix

\section{The decoding problem and the BP algorithm} \label{sect:background}

In a quantum fault-tolerant computing system, a syndrome extraction circuit periodically measures syndromes, and a decoder corrects errors based on the measurement results. Simulations of this process typically use one of two noise models. The circuit-level noise model, a realistic approach widely used in simulations, assumes every operation in the syndrome extraction circuit is a potential source of error. In contrast, the much simpler code capacity noise model only assumes idling noise on the data qubits, while all other operations are considered perfect. In this paper, we adopt the more realistic circuit-level noise model.

The decoding problem is formally defined by a parity-check matrix $\bfH\in \F_2^{M\times N}$, a logical operator matrix $\bfA \in \F_2^{K\times N}$, and a probability vector $\bfp=(p_0,\dots, p_{N-1})$. The $N$ columns of $\bfH$ and $\bfA$ correspond to the set of all possible error sources in the circuit-level noise model, with $p_i$ being the error probability of the $i$-th source. The $M$ rows of $\bfH$ correspond to the ``detectors", which are either syndrome measurement results or the XOR of two measurement results from consecutive rounds. An entry $\bfH_{ij} = 1$ if and only if the $j$-th error source flips the $i$-th detector. Similarly, the $K$ rows of $\bfA$ correspond to the logical operators, and $\bfA_{ij} = 1$ if and only if the $j$-th error source flips the $i$-th logical operator. The parity-check matrix $\bfH$ is also referred to as the detector error model in Stim~\cite{gidney2021stim}. Let $\bfe\in\F_2^N$ denote the binary vector of (unknown) error locations, and let $\bfs = \bfH\bfe \in \F_2^M$ be the observed syndrome. The decoder takes $\bfs$ as input and attempts to compute a correction $\hat{\bfe} \in \F_2^N$ such that it matches the syndrome ($\bfH\hat{\bfe} = \bfs$) and preserves the logical state ($\bfA\hat{\bfe} = \bfA\bfe$).

Both the new decoder proposed in this paper and our baseline for comparison, the BP-OSD decoder, are based on BP. We begin by reviewing BP decoding with the min-sum update rule. This algorithm operates on the Tanner graph, a bipartite graph constructed from the parity-check matrix $\bfH$. The graph's vertices are divided into $N$ error nodes ($e_0,\dots,e_{N-1}$), corresponding to the columns of $\bfH$, and $M$ detector nodes ($d_0,\dots,d_{M-1}$), corresponding to the rows of $\bfH$. An edge connects a detector node $d_i$ to an error node $e_j$ if and only if $\bfH_{ij}=1$. We use $\mathcal{N}(d_i)$ and $\mathcal{N}(e_j)$ to denote the set of neighbors for $d_i$ and $e_j$, respectively. The BP decoder functions by passing messages back and forth along these edges over multiple iterations. We use $D_{i\to j}(t)$ and $E_{j\to i}(t)$ to denote the detector-to-error and error-to-detector messages in the $t$-th iteration, respectively. The decoder takes the syndrome vector $\bfs=(s_0,\dots,s_{M-1})\in\F_2^M$ as input. To initialize, the decoder first calculates the prior log-likelihood ratio (LLR) for each error node as $\Lambda_j=\log\frac{1-p_j}{p_j}$. The initial error-to-detector messages in iteration 0 are then set to these priors: $E_{j\to i}(0)=\Lambda_j$ for all connected pairs $(d_i, e_j)$. In each subsequent iteration $t \ge 1$, the message passing proceeds in two steps. First, the detector-to-error messages are updated using the min-sum rule:
\begin{align}
D_{i\to j}(t) = (-1)^{s_i} & \cdot \prod_{e_{j'}\in\mathcal{N}(d_i)\setminus\{e_j\}} \sign(E_{j'\to i}(t-1)) \nonumber\\
& \cdot \min_{e_{j'}\in\mathcal{N}(d_i)\setminus\{e_j\}} |E_{j'\to i}(t-1)| . \label{eq:detector_to_error}
\end{align}
Second, the error-to-detector messages are updated by summing the prior LLR with all other incoming detector messages:
\begin{equation} \label{eq:error_to_detector}
E_{j\to i}(t) = \Lambda_j + \sum_{d_{i'}\in\mathcal{N}(e_j)\setminus\{d_i\}} D_{i'\to j}(t) .
\end{equation}
After updating the messages, the decoder also calculates the posterior LLR $\Lambda_j(t)$ for each error node as
\begin{equation} \label{eq:posterior_LLR}
\Lambda_j(t) = \Lambda_j + \sum_{d_{i'}\in\mathcal{N}(e_j)} D_{i'\to j}(t) .
\end{equation}
Based on this, a hard decision $\hat{e}_j(t)$ is made:
\begin{equation} \label{eq:hard_decision}
\hat{e}_j(t) = \left\{
\begin{array}{cc}
    0 & \text{if } \Lambda_j(t)> 0 ,  \\
    1 & \text{if } \Lambda_j(t)\le 0 .
\end{array}\right.  
\end{equation}
Define the vector $\hat{\bfe}(t) = (\hat{e}_0(t), \dots, \hat{e}_{N-1}(t))$. The iterative BP decoding stops at iteration $t$ if the syndrome equation $\bfH \hat{\bfe}(t)=\bfs$ is satisfied, or if $t$ exceeds a pre-fixed maximum number of iterations.

\begin{algorithm}
    \DontPrintSemicolon
    \SetAlgoLined
    \Func{\texttt{\emph{masked\_BP(edge\_msgs, pos\_val\_pairs, $\bfs$, max\_iters)}}}{
        \KwIn{a vector \texttt{edge\_msgs} used to initialize error-to-detector messages, a vector \texttt{pos\_val\_pairs} specifying the masked error nodes and their values, a syndrome $\bfs$, and maximum number of iterations \texttt{max\_iters}}
        \KwOut{A decoded error vector $\hat{\bfe}$ or declare failure\;}
        $\Lambda_j \gets \log\frac{1-p_j}{p_j}$ for $j=0,1,\dots,N-1$\;
        $[E_{j\to i}(0):(d_i,e_j) \text{ is connected}] \gets \texttt{edge\_msgs}$ \;
        \For{\emph{each $(j,v)$ pair in \texttt{pos\_val\_pairs}}}{
        \If{$v=1$}{Flip all $s_i$ such that $(d_i,e_j)$ are connected in the Tanner graph}
        }
        Error node $e_j$ is said to be masked if $(j,0)\in \texttt{pos\_val\_pairs}$ or $(j,1)\in \texttt{pos\_val\_pairs}$\;
        \For{\emph{$t=1,2,\dots,\texttt{max\_iters}$}}{
        Update messages and posterior LLRs according to \eqref{eq:detector_to_error}--\eqref{eq:posterior_LLR} but ignore all the messages to or from a masked error node. \;
        Calculate $\hat{e}_j(t)$ according to \eqref{eq:hard_decision} for all error nodes that are not masked. Fill the masked positions with $0$. \;
        \If{$\bfH \hat{\bfe}(t)=\bfs$}{
        \For{\emph{each $(j,v)$ pair in \texttt{pos\_val\_pairs}}}{
        $\hat{e}_j(t)\gets v$\;
        }
        \Return $\hat{\bfe}(t)$\;
        }
        } 
        \Return ``Failure"
    }
    \caption{Function \texttt{masked\_BP}}
    \label{algorithm:masked_BP}
\end{algorithm}

\begin{algorithm}
\DontPrintSemicolon
\SetAlgoLined
\KwParams{\texttt{max\_rounds}, \texttt{beam\_width}, \texttt{initial\_iters}, \texttt{iters\_per\_round}, \texttt{num\_results}}
\KwIn{A syndrome vector $\bfs$.\;}
\KwOut{A decoded error vector $\hat{\bfe}$.\;}
Initialize \texttt{results} as an empty set\;
Run BP for \texttt{initial\_iters} iterations and break out whenever $\bfH \hat{\bfe}(t)=\bfs$ is satisfied.\;
Insert the decoding result into \texttt{results} if BP succeeds, and return this result if \texttt{num\_results}=1.\;
\For{$j =0,1,\dots,N-1$}{
\texttt{sum\_LLR}$[j] \gets$  summation of $\Lambda_j(t)$ over all BP iterations.\;
}
Build an object \texttt{path} with 4 fields: \;
\texttt{path.edge\_msgs} $\gets [E_{j\to i}(t): d_i,e_j \text{ connect}]$, where $t$ is the index of the last BP iteration. \;
\texttt{path.pos\_val\_pairs} $\gets$ an empty vector \;
\texttt{path.next\_pos} $\gets \arg\min_{0\le j<N}|$\texttt{sum\_LLR}$[j]|$\;
\texttt{path.score} $\gets 0$ \;
Initialize \texttt{set} $\gets\{$\texttt{path}$\}$ with a single element \texttt{path}\;
\For{$r=1,2,\dots,$\emph{\texttt{max\_rounds}}}{
Initialize \texttt{next\_set} as an empty set\;
\For{\emph{each \texttt{path}$\in$\texttt{set} and \texttt{val}}$\in \{0,1\}$}{
Append $(\texttt{path.next\_pos},\texttt{val})$ to the end of \texttt{path.pos\_val\_pairs}\;
Run \texttt{masked\_BP(path.edge\_msgs, path.pos\_val\_pairs, $\bfs$, iters\_per\_round)} \;
If \texttt{masked\_BP} succeeds, insert the decoding result into \texttt{results}.\;
If size of \texttt{results} = \texttt{num\_results}, return the minimum weight element in \texttt{results}.\;
\texttt{iters} $\gets$ number of iterations actually run in \texttt{masked\_BP} (considering early return)\;
$\mathcal{U} \gets$ set of unmasked error nodes, i.e., $\{j:0\le j<N, (j,v)\notin \texttt{path.pos\_val\_pairs} \text{ for } v=0,1\}$, \;
\For{$j \in\mathcal{U}$}{
\texttt{sum\_LLR}$[j] \gets$  summation of $\Lambda_j(t)$ over all iterations in \texttt{masked\_BP}.\;
}
Build an object \texttt{nextp} with 4 fields: \;
\texttt{nextp.edge\_msgs} $\gets [E_{j\to i}(t): d_i,e_j \text{ connect}]$, where $t$ is the index of the last iteration in \texttt{masked\_BP} \;
\texttt{nextp.pos\_val\_pairs} $\gets$ \texttt{path.pos\_val\_pairs}\;
\texttt{nextp.next\_pos} $\gets \arg\min_{j\in\mathcal{U}} |$\texttt{sum\_LLR}$[j]|$\;
\texttt{nextp.score} $\gets \sum_{j\in\mathcal{U}}|\texttt{sum\_LLR}[j]|/\texttt{iters}$\;
Insert \texttt{nextp} into \texttt{next\_set}\;
Remove last entry of \texttt{path.pos\_val\_pairs}\;
}
Only keep the top \texttt{beam\_width} elements with the largest \texttt{score} in \texttt{next\_set} if its size exceeds \texttt{beam\_width}.\;
\texttt{set} $\gets$ \texttt{next\_set} \;
}
\caption{beam search decoder}
\label{algorithm:beam_search_decoder}
\end{algorithm}

\section{Detailed description of the beam search decoder} \label{sec:beam_search_decoder}

In this section, we present a detailed description of our new beam search decoder. Given the extensive notation required to define the algorithm, a summary of the relevant variables is provided in \cref{tab:meaning_of_variables} to assist readers.

The algorithm maintains a \texttt{set} of decoding \texttt{path}s. It is initialized by running some standard BP iterations to generate a single ``seed" \texttt{path}. The posterior LLRs from this initial run are used to identify the least reliable error node, which is stored in \texttt{path.next\_pos} and will be masked in future BP iterations.

\begin{table*}
\centering
\begin{tabular}{|l |p{10cm}|}
\hline
\textbf{Notation} & \textbf{Meaning} \\
\hline
\texttt{path} & An object representing a unique decoding path in the branching tree. \\
\texttt{path.pos\_val\_pairs} & An array containing indices and values of the error nodes that have been fixed for this path. \\
\texttt{path.edge\_msgs} & A snapshot of all error-to-detector messages from the last BP iteration, used to initialize the next round of masked BP for this path. \\
\texttt{path.score} & The path reliability score, used for sorting and pruning. \\
\texttt{path.next\_pos} & The index of the least reliable error node $e_j$ chosen as the branching point for the next round of masked BP iterations. \\
\texttt{nextp}   & The new child path object obtained by branching at \texttt{path.next\_pos} from the current \texttt{path} \\
\hline
\texttt{set} & The set of active candidate paths currently being processed. \\
\texttt{next\_set} & The set of candidate paths for the subsequent round, generated by branching every \texttt{path} in \texttt{set}. \\
\texttt{results} & A container for valid error vectors $\hat{\mathbf{e}}$ found (where $\mathbf{H}\hat{\mathbf{e}} = \mathbf{s}$). \\
\hline
\texttt{beam\_width} & The maximum number of candidate paths allowed in \texttt{set} at each step. \\
\texttt{max\_rounds} & The maximum depth of the branching tree (number of branching rounds) allowed. \\
\texttt{initial\_iters} & The number of BP iterations performed on the root node before the beam search begins. \\
\texttt{iters\_per\_round} & The number of masked BP iterations performed to update a specific path after an error node is fixed. \\
\texttt{num\_results} & The target number of valid solutions required to terminate the decoding process early. \\
\hline
\texttt{sum\_LLR[j]} & The cumulative posterior LLR for error node $e_j$. \\
\hline
\end{tabular}
\caption{Variables and notation used in the beam search decoder}
\label{tab:meaning_of_variables}
\end{table*}

The decoder then iterates through multiple rounds of ``masked BP". In each round, every \texttt{path} in the current \texttt{set} is expanded into two new branches by setting its \texttt{path.next\_pos} node to 0 and 1, respectively. Each new \texttt{path} inherits its parent's history of masked error nodes (stored in \texttt{path.pos\_val\_pairs}) and appends the new (node, value) pair to the history, using the node index from \texttt{path.next\_pos} and the value (0 or 1) just explored. A round of masked BP iterations is then run for each new \texttt{path}. The posterior LLRs in these BP iterations are used to determine two things: (1) the next least reliable unmasked node (which is stored as the new \texttt{path.next\_pos}), and (2) an overall reliability score for the \texttt{path}. This branching process doubles the size of \texttt{set} each round, so to maintain tractable complexity, the \texttt{set} is pruned back down to a fixed parameter \texttt{beam\_width} by keeping only the \texttt{path}s with the highest reliability scores.

We use $|\texttt{sum\_LLR}[j]|$ as a reliability metric for each error node $e_j$, where $\texttt{sum\_LLR}[j]$ is the summation of $\Lambda_j(t)$ over all BP iterations in the current round. The least reliable node to store in \texttt{path.next\_pos} is the one with the smallest $|\texttt{sum\_LLR}[j]|$. This sum-based metric is more robust than using the instantaneous magnitude $|\Lambda_j(t)|$ from the final iteration. While the instantaneous LLR reflects the decoder's confidence at that single step, it can be misleading. For example, an unreliable node whose posterior LLR oscillates between large positive and negative values would have a (correctly) low summed reliability, even if its final $|\Lambda_j(t)|$ is large.

The overall reliability score for each \texttt{path} is then calculated by summing the per-node reliability metrics, $|\texttt{sum\_LLR}[j]|$, over all unmasked error nodes and normalizing by the number of BP iterations in that round. This normalization is crucial because different \texttt{path}s may have run for a different number of iterations (due to early returns), and the division makes their scores comparable. In contrast, this normalization was not required when selecting the least reliable node (as described previously), since all nodes within a single \texttt{path} are guaranteed to go through the same number of iterations.

The beam search decoder has a parameter \texttt{num\_results} that controls its termination logic. If \texttt{num\_results} is set to 1, the decoder returns the first valid decoding solution it finds. Otherwise (for $\texttt{num\_results}>1$), the decoder maintains a set \texttt{results}, into which it adds every valid solution it discovers. As soon as this set reaches the target size of \texttt{num\_results}, the decoder terminates and returns the minimum-weight error vector from \texttt{results}, where the weight of an error vector $\hat{\bfe}=(\hat{e}_0,\dots,\hat{e}_{N-1})$ is defined as
$$
\wt(\hat{\bfe}) = \sum_{j=0}^{N-1} \hat{e}_j \log\frac{1-p_j}{p_j} \cdot
$$

Finally, a key optimization in our algorithm is the order of decoding different \texttt{path}s. In each round, the \texttt{path}s in the \texttt{set} are sorted by the reliability score and explored in descending order. We prioritize high-score \texttt{path}s first because they are intuitively more likely to lead to a correct decoding result.

A high-level sketch of the beam search decoder is presented in \cref{algorithm:sketch_beam_search} for the $\texttt{num\_results}=1$ case. We now provide a detailed explanation of the full algorithm, shown in \cref{algorithm:beam_search_decoder}. The decoder has five parameters: \texttt{max\_rounds}, \texttt{beam\_width}, \texttt{initial\_iters}, \texttt{iters\_per\_round}, and \texttt{num\_results}. We will describe the logic for the simplest case of $\texttt{num\_results}=1$.

The decoder begins by running \texttt{initial\_iters} standard BP iterations (following \eqref{eq:detector_to_error}–\eqref{eq:hard_decision}). If a valid solution is found (i.e., $\bfH \hat{\bfe}(t)=\bfs$), it is returned immediately. Otherwise, the decoder calculates a reliability metric $|\texttt{sum\_LLR}[j]|$ for every error node $e_j$, where $\texttt{sum\_LLR}[j]$ is the summation of $\Lambda_j(t)$ over all \texttt{initial\_iters} iterations.

This information is used to build the ``seed" \texttt{path} object, which initializes the \texttt{set} for the subsequent decoding rounds. This object contains four fields. The first, \texttt{path.edge\_msgs}, stores all error-to-detector messages from the last BP iteration, which will be used to ``warm-start" the next round of masked BP. The second, \texttt{path.pos\_val\_pairs}, stores the \texttt{(position, value)} pairs for all masked error nodes; for example, a pair $(j,v)\in\texttt{path.pos\_val\_pairs}$ means that error node $e_j$ is fixed to $v \in \{0,1\}$, and that $e_j$ is excluded from all the calculations in the masked BP. This field is initialized as an empty vector. The third field, \texttt{path.next\_pos}, stores the index of the next error node to mask, which is set to the node $e_j$ with the minimum reliability metric $|\texttt{sum\_LLR}[j]|$. The final field, \texttt{path.score}, is the \texttt{path}'s reliability score used for ranking. It is initialized to 0.

The decoder then executes up to \texttt{max\_rounds} rounds of masked BP. As described in the high-level overview, each round involves expanding the \texttt{set} by a factor of two and subsequently pruning it back to \texttt{beam\_width} based on \texttt{path.score} if its size exceeds this limit. We now provide a detailed explanation of the \texttt{masked\_BP} function, which is formally defined in \cref{algorithm:masked_BP}.

The \texttt{masked\_BP} function takes 4 input parameters. The first, \texttt{path.edge\_msgs}, "warm-starts" the BP run by initializing the error-to-detector messages from the previous round's final state. The second, \texttt{path.pos\_val\_pairs}, is a vector specifying the error nodes to be masked and their fixed values. The third is the original syndrome vector, $\bfs$. The final parameter, \texttt{iters\_per\_round}, specifies the maximum number of iterations for this run.
Before iterating, the decoder pre-processes the syndrome based on the values of masked nodes. For every masked error node $e_j$ that is set to $1$, all syndrome bits $s_i$ corresponding to detectors connected to $e_j$ are flipped. This transformation creates a modified syndrome, effectively recasting the decoding problem as one where all masked error nodes are fixed to $0$. 
The masked BP then runs for up to \texttt{iters\_per\_round} iterations, following the rules \eqref{eq:detector_to_error}–\eqref{eq:hard_decision} with a critical modification: all messages to or from masked error nodes are ignored. For example, in \eqref{eq:detector_to_error}, the message $D_{i\to j}(t)$ is not calculated if $e_j$ is masked; if $e_j$ is not masked, the calculation of $D_{i\to j}(t)$ will exclude any incoming messages $E_{j'\to i}(t-1)$ where $e_{j'}$ is masked.

Finally, we note a simplification in \cref{algorithm:beam_search_decoder} made for clarity and brevity. The pseudocode first gathers all new paths in \texttt{next\_set} (Line 28) and then applies a single, collective pruning step (Line 30). In our actual implementation, we maintain the set's bounded size ``on-the-fly" to improve efficiency.
A new path (\texttt{nextp}) is inserted into \texttt{next\_set} only if one of two conditions is met: (1) \texttt{next\_set} is not yet full (its size is less than \texttt{beam\_width}), or (2) \texttt{nextp.score} is greater than the minimum score currently in \texttt{next\_set}. In the second case, the path with the smallest score is ejected as \texttt{nextp} is inserted. This optimization is significant because it avoids the costly operation of copying \texttt{nextp.edge\_msgs} (a long vector) for paths that would be immediately discarded by the pruning step.

\end{document}